\documentclass[acmtog, nonacm]{acmart} 
\usepackage{hyperref} 
\AtBeginDocument{%
  \fancypagestyle{plain}{%
    \fancyhf{} %
    \fancyfoot[L]{ACM SIGENERGY Energy Informatics Review}%
    \fancyfoot[R]{Volume 5 Issue 3, September 2025}}
  \providecommand\BibTeX{{%
    \normalfont B\kern-0.5em{\scshape i\kern-0.25em b}\kern-0.8em\TeX}}}

\let\oldmaketitle\maketitle
\renewcommand{\maketitle}{%
  \oldmaketitle%
  \thispagestyle{plain}%
  \pagestyle{plain}}

\acmYear{2025}
\acmConference[DACH+]{Conference on Energy
Informatics 2025}{September 17--19,
  2025}{Aachen, Germany}

\usepackage{tikz}
\usetikzlibrary{decorations.pathreplacing}
\usetikzlibrary{arrows,shapes,positioning}
\usetikzlibrary{patterns}
\usetikzlibrary{positioning, fit, calc}

\setcitestyle{nosort,numbers,compress}

\newcommand\todo[1]{\textcolor{black}{#1}}

\begin{document}

\title[Processing of Grid Data for the Fair Spatial Distribution of PV Hosting Capacity]{Large-Scale Processing and Validation of Grid Data for Assessing the Fair Spatial Distribution of PV Hosting Capacity}

\author{Ali Mohamed Ali}
 \email{ali.alimohamed@hevs.ch}
 \affiliation{%
 \institution{HES-SO Valais-Wallis}
 \city{Sion}
 \country{Switzerland}}
 \author{Yaser Raeisi}
 \email{yaser.raeisi@hevs.ch}
 \affiliation{%
 \institution{HES-SO Valais-Wallis}
 \city{Sion}
 \country{Switzerland}}
 \author{Plouton Grammatikos}
 \email{plouton.grammatikos@hevs.ch}
 \affiliation{%
 \institution{HES-SO Valais-Wallis}
 \city{Sion}
 \country{Switzerland}}
 \author{Davide Pavanello}
 \email{davide.pavanello@hevs.ch}
 \affiliation{%
 \institution{HES-SO Valais-Wallis}
 \city{Sion}
 \country{Switzerland}}
 \author{Pierre Roduit}
 \email{pierre.roduit@hevs.ch}
 \affiliation{%
 \institution{HES-SO Valais-Wallis}
 \city{Sion}
 \country{Switzerland}}
 \author{Fabrizio Sossan}
 \email{fabrizio.sossan@hevs.ch}
\affiliation{%
 \institution{HES-SO Valais-Wallis}
 \city{Sion}
 \country{Switzerland}}

\begin{abstract}
The integration of PV systems and increased electrification levels present significant challenges to the traditional design and operation of distribution grids. This paper presents a methodology for extracting, validating, and adapting grid data from a distribution system operator's (DSO) database to facilitate large-scale grid studies, including load flow and optimal power flow analyses. The validation process combines rule-based sanity checks and offline automated power flow analyses to ensure data consistency and detect potential errors in the grid database, allowing for their correction. As a practical application, the paper proposes a method to assess the PV hosting capacity of distribution grids, with a focus on ensuring fairness in their spatial distribution. By incorporating fairness criteria into the analyses, we quantify the costs (in terms of missed revenues from selling PV generation) associated with spatial fairness.

\end{abstract}





\maketitle

\section{Introduction}
The integration of residential and commercial photovoltaic (PV) systems, along with the increasing electrification of demand through electric vehicles (EVs), challenges the conventional design and operational paradigms of distribution grids, which were originally designed as passive systems for one-directional power flows and well-defined levels of power demand at various nodes.

Adapting distribution grids to accommodate these increased levels of demand and generation requires distribution system operators (DSOs) to adopt new perspectives and take actions at multiple levels. The first line of action is generally analyzing and understanding whether their grids are capable of hosting prospective levels of distributed generation and demand. Because all distribution grids are different, this normally requires tailored analysis at the levels of individual feeders or networks by using deterministic or probabilistic load flow analysis to evaluate voltage levels at nodes, line currents, and power flows at transformers, and compare them with statutory levels and rated values.

The second step is planning appropriate countermeasures for those grids that might not be capable of accommodating increased levels of PV generation and demand. Countermeasures can be broadly classified into two kinds: traditional grid reinforcement (replacing or reinforcing power lines and substation transformers) or, as broadly advocated in the technical literature over the last two decades, alternative solutions such as flexible demand, energy storage systems, V1G or V2G. This classification is also known as wire and non-wire grid upgrades (e.g., \cite{10194742}). The choice between these two kinds of options is influenced by many factors, both internal and external to the DSO. These factors include investment costs, return on investment, public incentives, regulation aspects, the maturity and reliability of the technology, complexity, in-house availability of the required expertise to perform, operate, and maintain the upgrade, risk aversion, and innovation tendency. While traditional grid reinforcement is regarded as the safe choice for a DSO due to their mastery of the technology and techniques, alternative solutions face several adoption barriers due to being less established and requiring new competencies to design and operate. Despite this, they might offer better perspectives in terms of costs (especially compared to very extensive grid upgrades) and the provision of multiple services.


The third line of action involves developing and operating the identified countermeasures. The extent of operating these countermeasures can vary considerably depending on the selected solution. While traditional grid upgrades, once developed, require regular routine and non-routine maintenance, non-wire grid reinforcement necessitates the development of entirely new operational procedures. These include elements such as control and energy management systems, monitoring systems (e.g., control rooms, grid state estimation procedures), and services for forecasting power demand and PV generation. If non-wire grid reinforcement refers to utilizing assets not directly owned by the DSO but by third parties (e.g., aggregators, prosumers with behind-the-meter flexibility, V2G operators), appropriate contractual arrangements must also be established.

Regardless of the adopted grid upgrading strategy, a fundamental requirement for assessing the need for grid reinforcements and designing them is the availability of reliable and accurate data on distribution grids. Modern DSOs store their grid data (e.g., grid topology, line parameters, electrical equipment, and many other information) in relational databases. Given that a DSO, depending on the geographical area served, might easily have thousands or even tens of thousands of low- and medium-voltage nodes and lines, ensuring the accuracy of the stored information is certainly challenging. Spotting errors can be even more difficult. In the existing literature, bad-data detection techniques for power grid applications primarily refer to processing measurements from a monitoring system, with no focus on validating the actual grid data. For example, the works in \cite{aeiad2016bad, pignati2014pre} aim to identify erroneous data in order to exclude it from the grid-state estimation process; the study in \cite{pv_validation} assesses the level of accuracy required from monitoring equipment to ensure the reliable integration of PV generation into distribution grids. \todo{Similarly, more recent literature considers the development of methods for ensuring the reliability of measurements from smart meters for grid-state estimation in the phase of cyber attacks and false data injections (e.g., \cite{CHEN9706368,NAWAZ2021106819}), with no specific focus on using the measurements for correcting input grid data.}

In this context, this paper addresses the comprehensive process of extracting data from a real grid database of a DSO, followed by a validation step to ensure the consistency of the grid data, in particular the absence of gross errors. The validation step is twofold: first, a series of basic rule-based sanity checks on the raw database data is performed; second, offline power flows on whole ensembles of the grids are computed. Anomalies in the load flow results are then used to identify possible errors in the input data. This methodology was implemented in real-life and operated in collaboration with a DSO, proving beneficial for the identification and correction of errors in the dataset. Finally, as an application of the validated data, this paper proposes a method for evaluating the hosting capacity of the grid for additional PV installations, with a focus on the fairness of PV distribution across different nodes. Based on this analysis, we can assess the economic cost of achieving fair distribution and compare it against the cost of not having it.

In summary, the key contributions of this paper are the following:
\begin{itemize}
 \item An automatic toolchain to convert existing database data into actionable data for load flow analysis.
 \item A method for critically evaluating the integrity of the data to ensure the consistency of the grid data.
 \item The formulation of a problem to maximize the PV hosting capacity of a grid while accounting for spatial fairness.
\end{itemize}

The rest of this paper is organized as follows: Section~\ref{sec:methodology} details the methodology of the processing of the grid data used in this study, Section~\ref{sec:pv_allocation} presents the formulation of the optimization problem to maximize the PV hosting capacity in a given distribution grid and evaluates the impact of fairness on the total capacity. Finally, Section~\ref{sec:conclusion} concludes the paper by summarizing the key contributions.



\section{Methodology for grid data processing and validation}
\label{sec:methodology}

This section first provides a general overview of the utilization cycle of the grid data of a DSO, from stored data to applications. Then, it describes a methodology to extract grid data from a large database of a DSO, detect and correct data anomalies and errors, and finally put these data into a usable format for performing grid studies (load flows and optimal power flows).

\subsection{Grid data utilization cycle}
\begin{figure}[h!]
    \centering
    \includegraphics[width=0.95\linewidth]{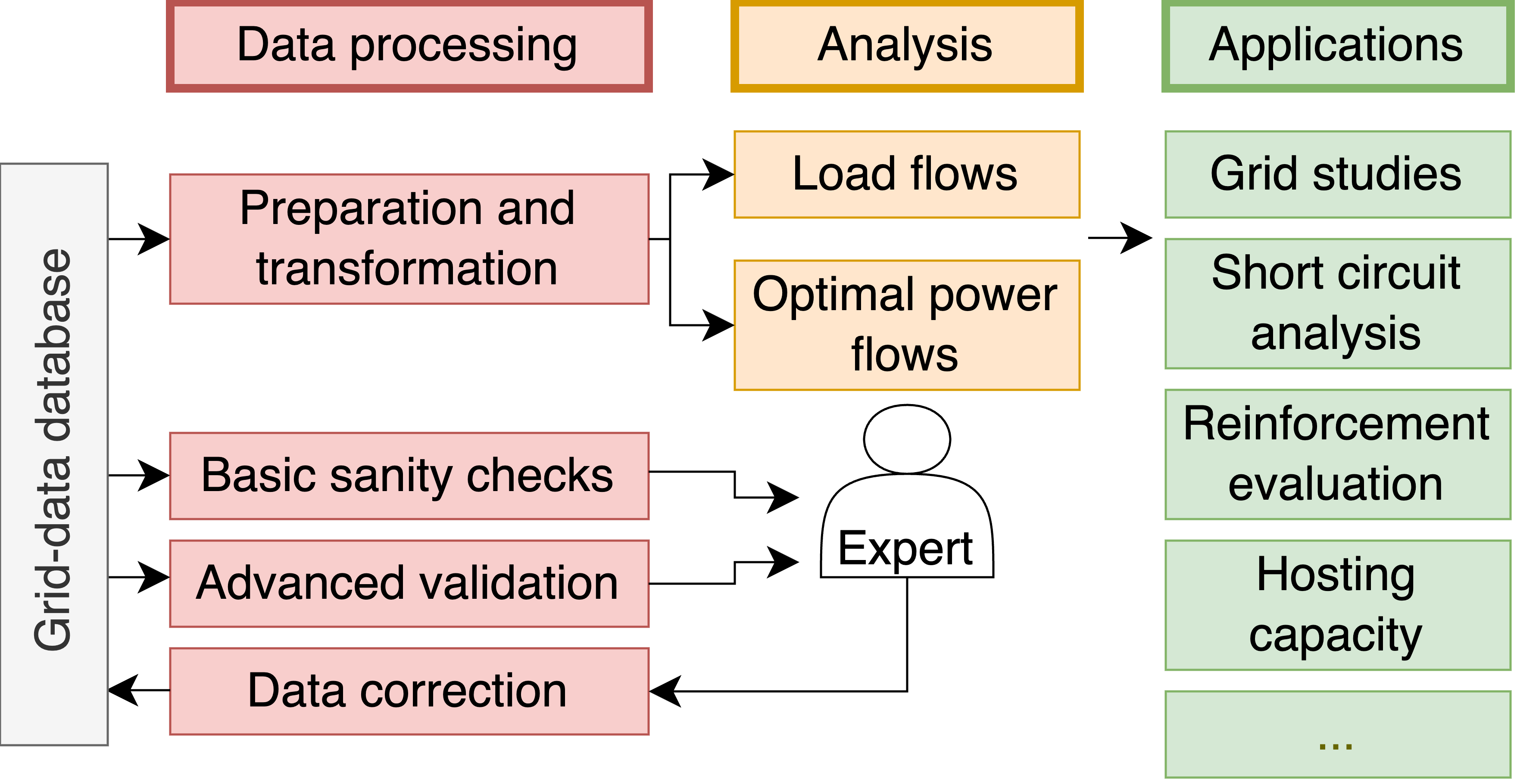}
    \caption{Grid data storing and utilization cycle.}
    \label{fig:overall workflow}
\end{figure}

Fig.~\ref{fig:overall workflow} illustrates the utilization cycle of DSO's grid data. From left to right, the first layer in the diagram represents grid data storage, typically implemented using a relational database. In this paper, grid data encompasses static electrical information of the grid, such as connectivity information, components, and their specifications. An example of this data is provided in the following subsection. The second layer in Fig.~\ref{fig:overall workflow} refers to data processing, which involves the preparation and transformation routines necessary to adapt the data for further analysis. These analyses include:
\begin{itemize}
    \item grid studies, which are typically performed using load flow or OPF models according to the requirements of the final applications (highlighted in green on the right side of the figure); and,
    \item the human-in-the-loop data correction procedure, where an expert operator from the DSO implements corrections to the dataset based on validation checks performed by automated data processing routines. 
\end{itemize}
It is worth highlighting that the second step mentioned above requires human expertise due to the extremely sensitive nature of modifying data in the database. As future work, the procedure proposed in this paper can be enhanced by developing more specific numerical methods to assist the operator even further. Integrating static data with other sources, such as measurements from the grid via smart meters, remote terminal units, or phasor measurement units, can provide additional support to the operator, for example by enabling the estimation of line impedances \cite{grammatikos2025measurement}.

\subsection{Types of grid data for distribution grid analyses}

The types of information that are typically necessary for distribution grid studies are the following:
\begin{itemize}
    \item List of substations (typically at multiple voltage levels, such as medium and low voltage), electrical cabinets, distribution boxes, junctions, and service entry boxes. These elements typically serve as the nodes of the grid.
    \item Component data within nodes: datasheet information about transformers and protective devices (circuit breakers, fuses, and switches), along with their operational states, such as the transformer tap positions and the switch status.
    \item Connectivity matrix highlighting connections among nodes.
    \item Lines between nodes: line type (overhead or buried), line length, electrical parameters (resistance, inductance, and capacitive susceptance) per unit of length, conductor type, and ampacity limit.
    \item Short-circuit impedance of the upper level grid.
    \item A list of generation resources and special loads connected at the grid buses.
\end{itemize}

\subsection{Data transformation for a commercial load flow solver}

We briefly describe data refactoring for processing grid information for enabling load flow calculations. As an example, we use DigSilent PowerFactory; however, the method can be adapted for other power solvers by changing the data model. This software uses the DGS interface for data exchange, supporting the import of full network models, updates to existing ones, and the export of model data and calculation results. The DGS data model is organized into three main categories: type data, element data, and graphic data, as described in the following.

Type data includes manufacturer-related information about components, e.g. the nominal current of components, the vector group and the load and no-load losses of transformers, and the line resistance and reactance (in ohm/km). 

Element data captures the operational attributes of components and the connectivity among them. Examples are the out-of-service status of components, the tap positions of transformers, and the line lengths in kilometers. 
Connectivity is defined as a list of entries specifying links between pairs of components.



Finally, the graphic data includes the graphic attributes of elements, such as position, size, or appearance.


This data refactoring procedure can be implemented as a script that reads information from the database and generates a DGS data structure as output. Once the data in DGS format is prepared and imported into PowerFactory, the grid operator can use its load flow analysis tool to perform grid studies and its graphical interface to visualize and interpret results.



\subsection{Grid data validation}
This step is essential to ensure that the data extracted from the database is valid and that analyses based on it are trustworthy. We implement two types of validation and sanity checks: basic and advanced, as described below. Both methods output messages for the grid expert to facilitate the identification of errors. Because databases of grid operators might include tens of thousands of entries, future work can focus on implementing more sophisticated data identification and correction strategies to further assist the data correction procedure \todo{such as, corrections in geographic data which is either incorrect or doesn't show the actual connection}.

\subsubsection{Basic validation}
Basic validation is implemented by rule-based checks to identify trivial errors in the data. These include:
\begin{itemize}
\item Invalid topologies in the network or within nodes. Examples include explicit meshed configurations in LV grids appearing in low-voltage distribution grids or fuses in parallel.
\item GPS coordinates of the nodes outside the operating area of the operator.
\item Cable length between two nodes significantly larger than the Manhattan distance between them.
\item Cable sections that are too small or too large. 
\item Missing attributes of the components (e.g., rating).
\end{itemize}

\subsubsection{Advanced data validation through offline load flow computations}
\todo{
As an original contribution of this paper, we propose a validation strategy aimed at detecting gross inconsistencies in the grid database. The procedure consists of computing offline multi-period load flows under nominal grid conditions and identifying electrical quantities that exceed the statutory or physical operating limits of the distribution grid. The underlying principle is that, if the stored grid data is accurate, such violations should not occur. In contrast to the basic validation strategies described above, which focus on targeted analyses of individual attributes, this approach can be regarded as a more global attempt to validate the overall consistency of the data. It is worth highlighting, however, that the procedure merely indicates the presence of inconsistencies, providing the operator with the opportunity to double-check the data. More specific efforts, such as attempting to pinpoint the exact element causing the inconsistency, will be developed as part of future work.}

\begin{figure}[ht]
\centering {
\footnotesize
\sffamily
\tikzstyle{block} = [rectangle, draw, fill=gray!10, 
    text width=20em, text centered, rounded corners, minimum height=3em]
\tikzstyle{line} = [draw, -latex']

\begin{tikzpicture}[node distance = 0.95cm, auto]
    \node [block] (0) {Retrieve node connectivity information from the database};
    \node [block, below of=0] (1) {Build the branch-to-node matrix};
    \node [block, below of=1] (2) {Retrieve lines attributes};
    \node [block, below of=2] (3) {Calculate primitive admittance matrix};
    \node [block, below of=3] (4) {Calculate bus admittance matrix};
    \node [block, below of=4] (5) {Retrieve load information};
    \node [block, below of=5] (6) {Calculate load curves over time from models};
    \node [block, below of=6] (7) {Compute multi-period load flows};
    \node [block, below of=7] (8) {Calculate criteria for the identification of anomalous load flow results};
    
    \path [line] (0) -- (1);
    \path [line] (1) -- (2);
    \path [line] (2) -- (3);
    \path [line] (3) -- (4);
    \path [line] (4) -- (5);
    \path [line] (5) -- (6);
    \path [line] (6) -- (7);
    \path [line] (7) -- (8);
\end{tikzpicture}
}
\caption{Procedure for automatic validation based on load flow results.}\label{fig:flow_diagram}
\end{figure}
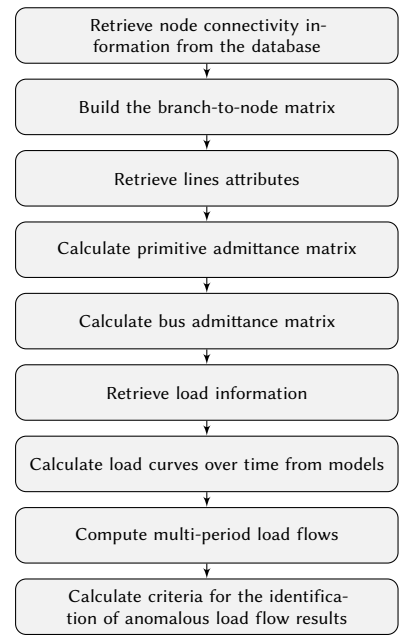

The process is summarized in Fig.~\ref{fig:flow_diagram} and explained briefly hereafter. The information on connectivity among nodes is retrieved from the database, along with the electrical and physical attributes of the lines, and used to build the bus admittance matrix by combining the well-known notions of line-to-node matrix and primitive admittance matrix \cite{grainger1999power}. Then, time series data for power demand and distributed generation are either collected from measurements or, if unavailable, synthesized using per-unit load profiles scaled to match the nominal demand and generation at each node (e.g., \cite{CIGREREF}). Based on this data, a single-phase equivalent load flow analysis is performed. A three-phase load flow is currently considered unnecessary for the purpose of data validation. 
Anomalous load flow results are identified based on the following criteria:
\begin{itemize}
    \item Voltage magnitudes outside the statutory limits (i.e., ±10\% in low-voltage networks and ±5\% in medium-voltage networks);
    \item Current magnitudes exceeding the ampacity limits of lines or cables;
    \item Current magnitudes exceeding the ampacity limit at the grid connection point.
\end{itemize}

\todo{
\subsubsection{Application and results of the method}
The proposed validation method was applied to the data provided by a local DSO of a network consisting of 1,100 MV/LV substations. First, the basic validation step led to the identification of several errors in 62\% of the networks in the DSO's database, including missing cable lengths and non-standard cable sections. The errors were corrected using our rule-based sanity-check methodology, such as by replacing missing cable sections with standard values from datasheets and filling in missing cable lengths with a fixed value. This step also revealed that 13\% of networks had some topology configuration problem (e.g., a meshed configuration). Then, the load flow analysis performed during the advanced validation step revealed that 36\% of the networks violate the nodal voltage limits or the ampacity limits of lines or cables. These networks, together with the ones with configuration problems, were communicated to the DSO for further investigation. Finally, the networks that passed both validation procedures can be used for various grid studies. One such example is the fair allocation of PV hosting capacity, which is presented in the following section.
}


\section{Fair allocation of PV hosting capacity in a distribution grid}
\label{sec:pv_allocation}

As an application of the grid data processing presented in the previous section, we consider the use case of calculating a fair distribution of PV hosting capacity on a given distribution grid, accounting for its operational constraints. The notion of fairness in power systems is a broad subject that deals with ensuring equal access to the electrical grid infrastructure regardless of factors such as the grid connection point or the wealth of the household. Several definitions and metrics have been used in the literature to quantify fairness in the context of congestion management and electricity markets (e.g., \cite{evaluating_fairness, 10863079}), with applications that include both planning and operational aspects, such as control of resources, PV curtailment, and network reconfiguration (\cite{item_0215aa913aca4656b29e3e92bfc0b0de,8260294, GEBBRAN2021116546,de2024review}).


In this section, we examine the fairness of allocating PV generation capacity within distribution grids. Ideally, all prosumers should have the opportunity to connect their desired amount of PV capacity, regardless of their location in the grid and without their decision being influenced by the decisions or actions of other users; however, this might not be possible due to physical limits of the distribution grid infrastructure. In particular, fairness issues can be manifested in two ways:
\begin{itemize}
    \item Location: grid connection points closer to a substation transformer will have access to higher power capacity due to the smaller impedance of the cables and lines (spatial fairness \cite{8626512, WANG2025101648}); and
    \item Timeline of adoption: early adopters of PV generation, such as wealthy households, might saturate the grid's PV generation capacity, delaying the connection of PV generation for later adopters due to the need for grid reinforcement.
\end{itemize}

In this context, the objective of this section is to develop a method that enables a DSO to (i) quantify the grid's capacity to host PV generation, considering both location (i.e., installation site) and installed power, and (ii) allocate this capacity fairly, ensuring equal access to PV generation among the connected prosumers (spatial fairness). We will show that incorporating spatial fairness comes at the cost of reducing the total hosting capacity of the grid. As will be discussed, a virtue of the proposed approach is its ability to quantify this cost of fairness by comparing it to the lost revenue resulting from reduced electricity PV production.





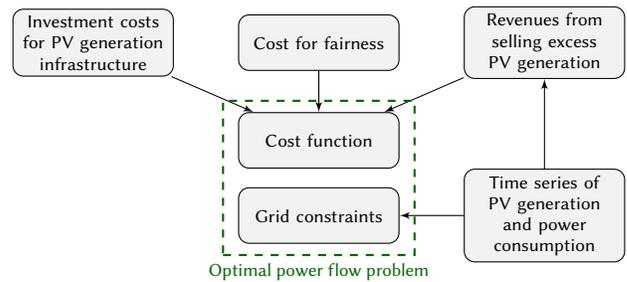
\begin{figure}[ht]
\centering {
\footnotesize
\sffamily
\tikzstyle{block} = [rectangle, draw, fill=gray!10, 
    text width=8em, text centered, rounded corners, minimum height=3em]
\tikzstyle{line} = [draw, -latex']

\begin{tikzpicture}[node distance=1.3cm, auto]
    \node [block] (0) {Cost function};
    \node [block, above of=0, xshift=-3.0cm] (1) {Investment costs for PV generation infrastructure};
    \node [block, above of=0, xshift=3.0cm] (2) {Revenues from \\ selling excess PV generation};
    \node [block, above of=0, xshift=0.0cm] (3) {Cost for fairness};

    \node [block, below of=0, node distance=1cm] (constraints) {Grid constraints};
    \node [block, below of=2, node distance=2.3cm] (pv) {Time series of PV generation and power consumption};
    
    \path [line] (1) -- (0);
    \path [line] (2) -- (0);
    \path [line] (3) -- (0);

    \path [line] (pv) -- (2);
    \path [line] (pv) -- (constraints);

\node[draw, green!40!black, thick, dashed, inner xsep=2mm,inner ysep=1.5mm,fit=(0)(constraints)](box){};
\node[below of=constraints, color=green!40!black, node distance=0.75cm] {Optimal power flow problem};

\end{tikzpicture}
}
\caption{Overview of the formulation of the PV hosting capacity problem with fairness.}\label{fig:problem_formulation}
\end{figure}

Fig.~\ref{fig:problem_formulation} illustrates the proposed problem formulation, which will be detailed in this section. In a nutshell, it consists of an optimal power flow (OPF) problem that aims to minimize the total economic cost \todo{of the prosumers}, comprising capital investment for PV generation minus revenues from selling excess PV generation to electricity markets and augmented by a non-economic fairness term, for which we offer an economic interpretation through a Pareto front analysis. The problem is subject to grid constraints, including nodal voltages, line currents, and power at the substation transformer, and is driven by time series data of power consumption and PV generation at the grid nodes.

\todo{It should be noted that the economic and fairness objectives are formulated from the perspective of the prosumers, while the DSO, in vertically unbundled power grids, is concerned only with grid constraints. An interesting direction for future work would be to incorporate the economic benefits of the DSO (i.e., revenues from grid tariffs) into the cost function. It is also worth highlighting that our focus is on grid capacity and fairness.}

\todo{PV generation potential is an input to the problem and can be incorporated as input from existing standard tools considering land-use constraints and roof-top potential (e.g., \cite{assouline2018large, sossan2016large}).}

\subsection{Calculation of the maximum PV hosting capacity}
We begin by addressing the maximization of the PV hosting capacity of the distribution grid within its operational and physical limits, with no regard to fairness at this stage. The objective is to compute the generation capacity of distributed PV power plants (kWp of each plant) at each node of the grid that:
\begin{itemize}
    \item maximizes economic performance, given by the revenues from selling PV electricity at given electricity market prices and the total investment for the selected PV generation infrastructure;
    \item while subject to the constraints of the electrical infrastructure without any grid upgrades.
\end{itemize}

The notation is as follows: the index $n = 1, \dots, N$ denotes the $n$-th node, among a total of $N$ nodes in the grid, and $\alpha_n$ represents the PV generation capacity (in kWp) installed at node $n$. The indices $t = 1, \dots, T$ correspond to a discretized time interval of duration $\Delta$ (in hours), with $\Delta = 1/6$ hours (10 minutes), over a total of $T$ intervals.


\subsubsection{PV generation model}
\begin{figure}
    \centering
    \includegraphics[width=\linewidth]{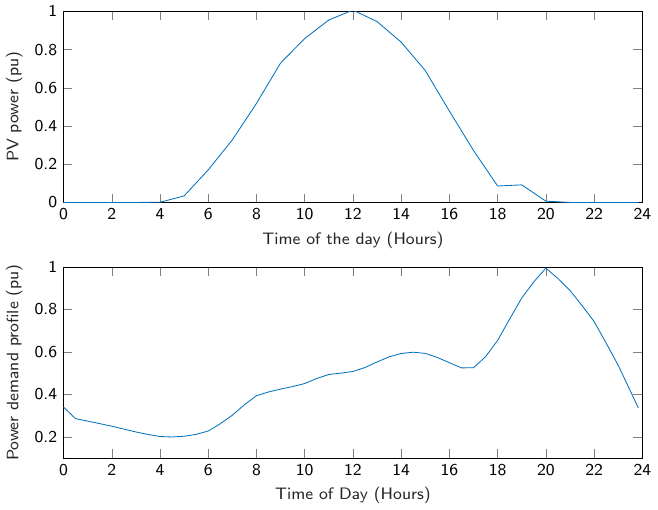}
    \caption{Nominal PV generation (top panel) and power consumption (bottom panel) throughout the day.}
    \label{fig:PV_and_load_profiles}
\end{figure}

PV production depends on the size of the installation, the plane-of-array irradiance on the panels, their temperature, operational status, and efficiency, as well as the efficiency of the converter. In power system studies, it is typically represented using simplified models that nonetheless account for these dependencies (e.g., \cite{friesen2018photovoltaic, sossan2019solar}).
For simplicity, and due to the comparative nature of the analysis proposed in this paper, we assume that PV installations have a uniform layout across the network (i.e., tilt and azimuth are optimized for maximum annual yield), with no differences in shading conditions.
Additionally, irradiance and temperature conditions are assumed to be uniform across all panels, owing to the spatial proximity of the installations (being located within the same grid) and the coarse temporal resolution of the analysis (10-minute intervals). The only differentiating factor, due to being the unknowns of the problem, is the considered installed capacity $\alpha_n$ across the nodes. In these settings, the distributed PV generation, denoted by $P_{n,t}^{PV}$ to represent the average PV generation at node $n$ in the time interval $t$ (in kW), is modeled as:
\begin{align}\label{eq:pv_max}
    P^{PV}_{n,t} = \alpha_n \cdot \widehat{G}_t
\end{align}
where $\widehat{G}_t$ is the production profile of a nominal 1~kWp PV plant, and it is shown in Fig.~\ref{fig:PV_and_load_profiles}.


\subsubsection{Cost function}
We aim to compute the installed PV generation capacity at the various nodes of the grid $\alpha_1,...,\alpha_N$, that we collect into the vector $\alpha = [\alpha_1, \dots,\alpha_N]$ for convenience in the formulation. The problem's cost function includes the following terms.
\begin{itemize}
    \item The investment cost for installing PV plants across the network:
    \begin{align}\label{eq: CAPEX}
    \mathcal{J}^{C}(\alpha) = \sum_{n=1}^{N} \alpha_n \cdot c^{cap},
    \end{align}
    where $c^{cap}$ is the turnkey unit costs of the installed PV capacity in CHF/kWp. 
    \item The operational costs associated with selling excess (non-self-consumed) PV generation to the grid. We consider the perspective of a (aggregated) prosumer at node $n$ with behind-the-meter PV generation. Its net power consumption is the difference between load consumption $P^{L}_{n,t}$ and PV production, thus $P^{L}_{n,t}-P^{PV}_{n,t}$: the positive part $[\cdot]^+$ of this term is the residual consumption, supplied by the grid, and priced at the retail electricity tariff $c^{+}$ (CHF/kWh); its negative part $[\cdot]^-$ is the feed-in to the grid, compensated with a tariff $c^{-}$ (CHF/kWh). Generally, $c^{+} \le c^{-}$ due to grid utilization tariffs and balancing costs. The total electricity bill of the $N$ prosumers over a period of $T$ time intervals can be calculated as: 
\begin{align}\label{eq: OPEX1}
     \mathcal{J}_0^{O}(\alpha) = \sum_{n=1}^{N} \sum_{t=1}^{T} \left( c^+ \left[P^{L}_{n,t}-\alpha_n \widehat{G}_t\right]^+  
 - c^{-}\left[P^{L}_{n,t}-\alpha_n \widehat{G}_t\right]^- \right) \Delta.
\end{align}
\end{itemize}

The cost function above is the difference of two convex functions and thus is nonconvex. However, it can be reformulated into a convex cost function, as described in \cite{GUPTA2023120955}, which is the formulation we ultimately adopted.

The operational cost should be adjusted for the considered lifetime, $r$ (in years), of the project, and the future expected revenues for the present value. For this, we use the net present value (NPV) concept, considering an annual discount rate $i$ (\%). The closed-form expression of the NPV factor used is:
\begin{align}\label{eq: NPV formula}
    NPV = \dfrac{1-(1+i)^{-r}}{i}.
\end{align}
Therefore, considering the net present value of future cashflow results in a reformulation of the operational costs in Eq. \eqref{eq: OPEX1} as follows:
\begin{align}
    \mathcal{J}^{O}(\alpha) = \mathcal{J}^{O}_0(\alpha) \cdot NPV
\end{align}

\subsubsection{Grid constraints}
The constraints of the optimization problem refer to respecting the physical and statutory (static) limits of the distribution grid. These include maintaining voltage magnitudes at the nodes within statutory bounds, ensuring line currents remain below ampacity limits, and keeping the active and reactive power at the substation transformer within its apparent power rating.
Grid constraints are formulated using a linearized load flow model based on sensitivity coefficients, derived from grid injection equations, following the approach in \cite{ChristakouLinearModel}. This method is widely adopted in the literature, and its description is omitted here for brevity.

While the reactive power of the load is considered in the problem in the input power demand time series, PV plants are assumed to operate at a unitary power factor (i.e., zero reactive power). Although reactive power support from PV inverters could be incorporated into the formulation, it is not of particular interest in this low-voltage grid use case due to the predominantly resistive nature of the distribution lines.

\subsection{Inclusion of fairness into the problem}
The former problem computes the maximum installable PV generation capacity in the network, ensuring full utilization of the grid’s capacity within its physical constraints. However, this will lead to spatial inequalities across installed PV generation, such as nodes located closer to the grid connection point might accommodate more PV generation due to shorter line lengths (and reduced voltage variations). To implement fairness, we augment the cost function with a penalty term that discourages configurations with unequal spatial distribution of PV generation. As a measure of spatial inequality (thus of spatial unfairness), we use the variance of the (rescaled) installed PV generation capacities across the network nodes. Let $\overline{p}_n$ denote the nominal power of node $n$, so that the ratio $\alpha_n / \overline{p}_n$ represents the installed per-unit PV generation capacity, relative to the node's nominal power. The measure of spatial unfairness is thus defined as the variance of the per-unit PV installed capacity. Formally, it is:
\begin{align}\label{eq:cost_fair}
 \mathcal{M}^{\text{U}}(\alpha) = \dfrac{1}{N-1}\sum_{n=1}^{N} \left (\dfrac{\alpha_n}{\overline{p}_n}-   \overline{\left(\dfrac{\alpha}{\overline p}\right)}\right)^2,
\end{align}
where
\begin{align}
    \overline{\left(\dfrac{\alpha}{\overline{p}}\right)} = 
    \frac{1}{N} \sum_{n=1}^N \dfrac{\alpha_n}{\overline p_n}
\end{align}
reads as the average of the per-unit PV installed capacity across the grid. Equation~\eqref{eq:cost_fair} yields a high value when the distribution of PV generation is spatially uneven across the network, and zero when PV plants are uniformly distributed (i.e., when the variance is zero). Notably, this cost term is convex, as it results from the composition of a convex function with a linear function. Therefore, incorporating this term into any cost function preserves the convexity of the original optimization problem. 

The final formulation of the optimization problem for fair PV hosting capacity allocation is
\begin{subequations}
\label{eq:optimproblem}
\begin{align}
\min_{\alpha} \;\; \left\{
\mathcal{J}^{C}(\alpha) + 
\mathcal{J}^O(\alpha) +
\lambda \cdot \mathcal{M}^{U}(\alpha)
\right\} \label{eq: cost function}
\end{align}
subject to:
\begin{align}
&\begin{aligned}
&\text{Grid constraints for all nodes, lines, and} \\
&\text{~~time intervals, and substation transformer;}
\end{aligned}\\
&\begin{aligned}
& \text{Nodes' nominal powers:}\\
& ~~-\overline{p}_n \leq P^L_{n,t}-\alpha_n \cdot \widehat{G}_t \leq \overline{p}_n & \; \forall n,t.
\end{aligned}
\end{align}
\end{subequations}
where the weight $\lambda$ (in CHF/pu$^2$) can be interpreted, in an economic way, as the tariff that penalizes unfair (unequal) spatial distribution of PV generation. Large values of $\lambda$ promote fairness, and vice versa. Evaluating the term $\lambda \cdot \mathcal{M}^U(\cdot)$ at the optimal solution $\alpha^\star$ provides a monetary quantification of the cost attributed to fairness. An example illustrating this interpretation will be described in the Results section.

\subsection{Case study network}
\begin{figure}[!ht]
    \centering
    \includegraphics[width=\linewidth]{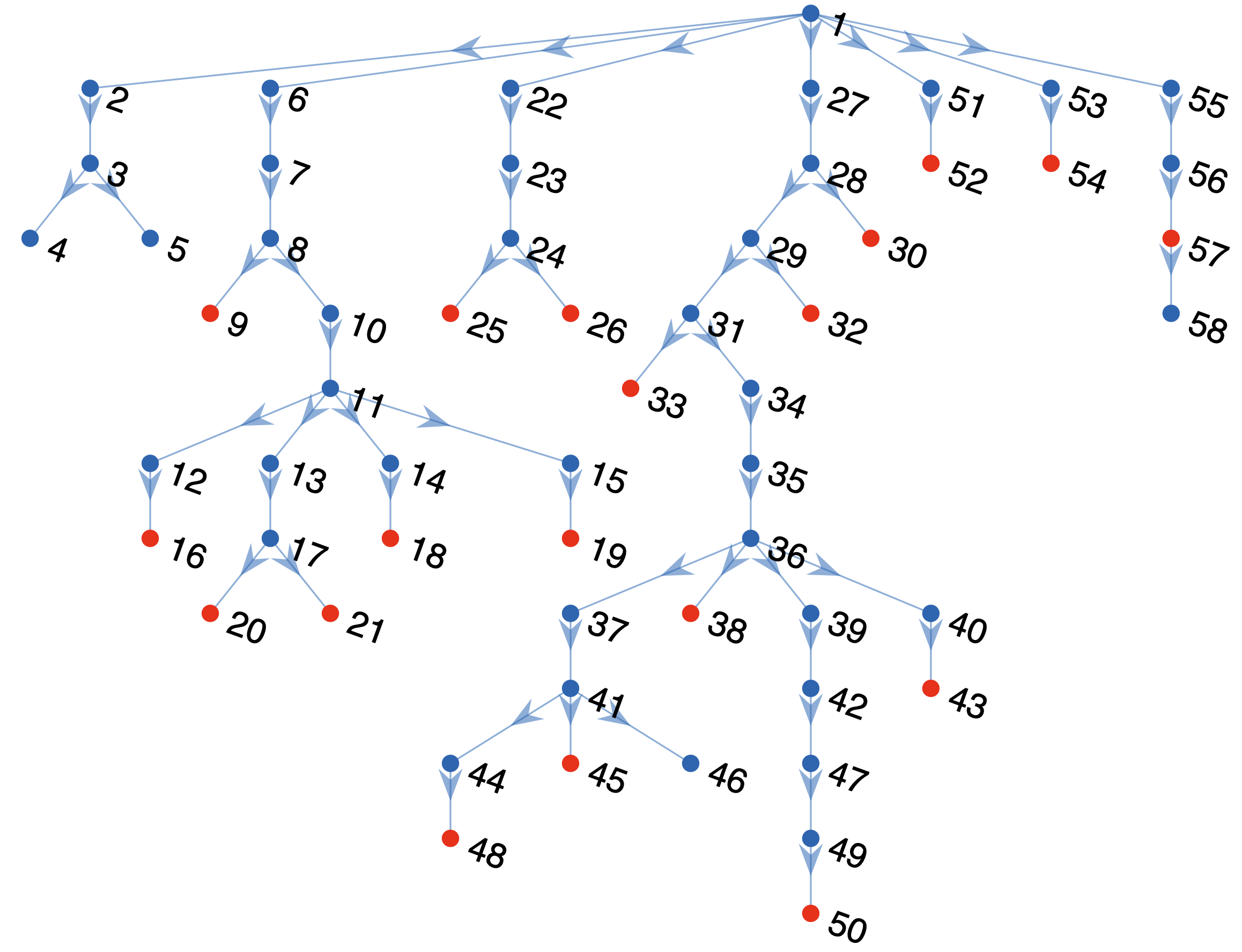}
    \caption{Topology of the case study network. Red nodes represent load connections, while blue nodes indicate junctions or potential future grid extensions.}
    \label{fig:network topology}
\end{figure}

\begin{figure}[!ht]
    \centering
    \includegraphics[width=\linewidth]{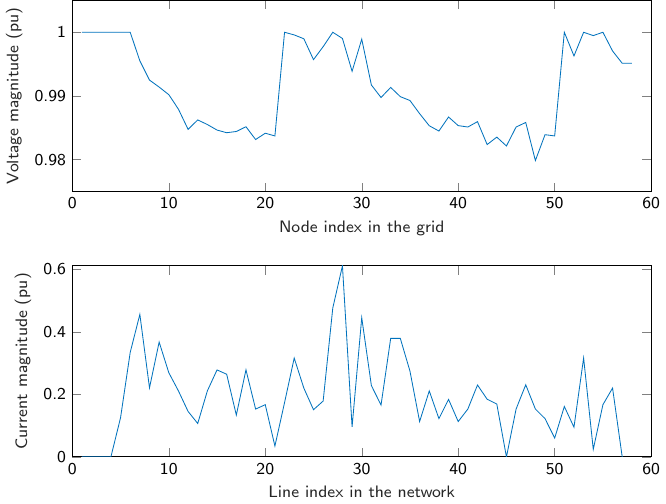}
    \caption{Nodal voltages (top panel) and line currents (bottom panel) calculated by the single-phase equivalent single-period load flow model of the considered network.}
    \label{fig:snapshot_loadflow}
\end{figure}

\todo{While the method is general and can be applied to any grid for which data is available, we apply it to a real-world case study to evaluate its performance and draw conclusions on the hosting capacity of an actual system.}
We consider a real-world three-phase low-voltage (400~V) network with a radial topology, as shown Fig.~\ref{fig:network topology}. The network has seven feeders, a total of 58 nodes, and 19 nodes with electrical loads, identified by red dots in Fig.~\ref{fig:network topology}. The network hosts a total demand of 319~kW, assumed at 0.9 power factor (154 kVAR), supplied by a transformer rated at 630~kVA. Fig.~\ref{fig:snapshot_loadflow} shows the voltage magnitudes across the network, calculated using a load flow analysis in a single-phase equivalent model, under conditions of maximum power demand at all nodes and slack voltage of 1 pu. Because the voltage magnitudes remain well within the $\pm 10\%$ of their nominal per unit value, and the currents are within acceptable limits, the results indicate a well-sized network with possibly no errors in the topological and electrical input data. Therefore, the PV hosting analysis can proceed.

The PV hosting analysis procedure requires evaluating power injections throughout the day due to the variable nature of the PV production and power demand. Because time series of the power demand at low voltage were not available from the operators, we use the synthetic load consumption profile from \cite{CIGREREF} and shown in the bottom panel of Fig.~\ref{fig:PV_and_load_profiles}, rescaled according to the nominal power consumption of each node. 

\subsection{PV economic parameters and tariff assumptions}

\begin{table}
\centering
\begin{tabular}{lccr}
\toprule
 Parameter & Symbol & Unit & Value  \\
 \midrule
 PV unit cost & $c^{cap}$ & CHF/kWp & 1'500 \\
 PV plant lifespan & $r$ &years & 20\\
 Discount rate & $i$ & \% & 2\\
 Retail electricity tariff & $c^+$ & CHF/kWh & 0.25\\
 Feed-in tariff & $c^-$ & CHF/kWh & 0.14\\
 \bottomrule 
\end{tabular}
\caption{PV economic parameters and tariff assumptions.}
\label{tab:economic parameters}
\end{table}
Table \ref{tab:economic parameters} presents the economic parameters used as input in this work. The PV unitary cost and discount rate reflect the current market in Switzerland, while the PV plant lifespan considered is based on existing literature. In addition, the retail electricity tariff is from the DSO of the case study network, and the feed-in tariff is the average of peak and off-peak prices of the DSO (This is done for simplicity, as this work is focused on technical constraints rather than the dynamics of electricity markets). 

\subsection{Results}
This section presents the result of the PV hosting maximization considering investment and operational costs, as well as the fairness of PV distribution.

\subsubsection{Fair and unfair allocation of PV hosting capacity}
\begin{figure}[!ht]
    \centering
    \includegraphics[width=\linewidth]{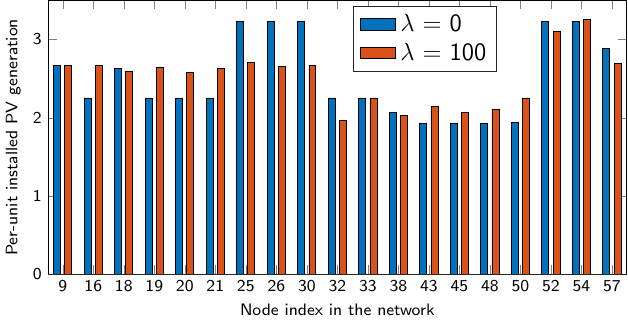}
    \caption{Allocated PV hosting capacity per node normalized by the nominal power of the node $\overline{p}_n$ for two values of the fairness weight $\lambda$.}
    \label{fig:fairness_allocation}
\end{figure}

To evaluate the impact of fairness on the geographical distribution of installed PV generation capacity, we solve Problem~\eqref{eq:optimproblem} for different values of $\lambda$. Fig.~\ref{fig:fairness_allocation} shows the installed PV generation capacity across the grid nodes (rescaled by each node nominal power) for two example values of $\lambda$: 0 and 100. We can see that when a higher emphasis is placed on fairness (higher $\lambda$), the algorithm reallocates PV capacity from nodes close to the feeder, e.g, nodes 25 and 26, to those located further away, e.g, nodes 20 and 21, achieving a more uniform distribution of installed capacity.

\subsubsection{The price for fair spatial allocation of PV generation capacity}

\begin{figure}
    \centering
    \includegraphics[width=\linewidth]{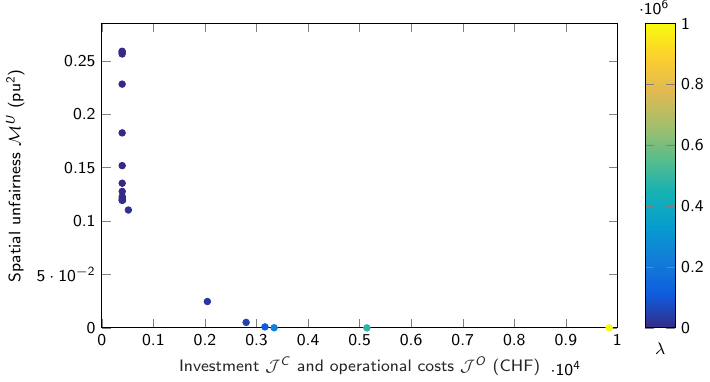}
    \caption{Spatial unfairness against total expenditure (investment and operational costs) of PV hosting for different values of $\lambda$.}
    \label{fig:fairness versus totalcost}
\end{figure}

Fig.~\ref{fig:fairness versus totalcost} shows the unfairness metric $\mathcal{M}^U$ versus the monetary cost associated with PV generation investment and operation, namely $\mathcal{J}^C +\mathcal{J}^{O}$, evaluated at the optimal hosting capacities for a range of values of $\lambda$. We observe that for smaller values of $\lambda$, it is possible to double the fairness (i.e., halve the variance) of PV hosting capacity in the network by reducing the profit by less than 1\% over the given 20-year lifespan. On the other hand, if very high degrees of fairness are required, one should pay, for this grid use case, several thousand CHF more to achieve only marginal improvements in variance.

Therefore, fairness in PV hosting capacity is inherently dependent on the specific objectives and priorities set by the decision maker. For example, they could specify the budget they are willing to sacrifice for fairness and choose the weight $\lambda$ accordingly. Nevertheless, this analysis allows us to quantify the cost the system has to sustain to give equal opportunities to everyone on the network to host PV capacity regardless of their location.

\subsubsection{Impact of fairness on the PV hosting capacity}
\begin{figure}
    \centering
    \includegraphics[width=\linewidth]{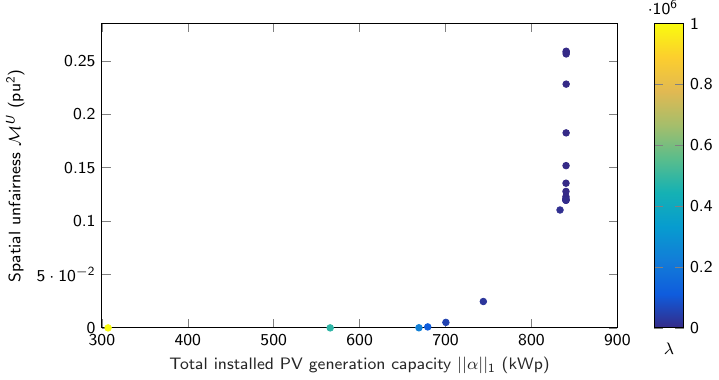}
    \caption{Spatial unfairness against total installed PV generation capacity for different values of $\lambda$.}
    \label{fig:fairness vs total capacity}
\end{figure}

Finally, we explore the impact of increasing fairness on the total PV generation capacity installed in the network. The results are shown in Fig.~\ref{fig:fairness vs total capacity} for different values of $\lambda$. First, one can observe that if no focus is given to spatial fairness (i.e., largest value of variance), one can install approximately 840 kWp of PV generation capacity in this grid. Then, one can halve the variance of PV installation among the nodes by a factor of two (from approx. 0.25 to 0.1) by giving up a few kWp of installed capacity, a relatively small price to pay for a more uniform distribution of PV generation capacity across the network. 

However, from Fig.~\ref{fig:fairness vs total capacity}, the total installed capacity may decrease from 840 kWp to 300 kWp if ideal fairness is enforced. This is because perfect fairness requires allocating equal generation capacity to all nodes. Consequently, the node with the lowest installed capacity, the weakest grid node, sets the upper limit for all others. This leads to a highly conservative allocation of PV generation capacity from the perspective of grid constraints. Therefore, it is important to carefully evaluate the trade-off between fairness and total PV capacity to avoid unnecessarily limiting potential capacity in pursuit of fairness.

\todo{It is worth highlighting that these results are specific to the investigated grid use case. However, the proposed method is general and will be applied to a larger dataset from the operator to draw conclusions that can be extended to the operator’s entire operational area.}

\section{Conclusions}
\label{sec:conclusion}
We have presented a methodology for processing and validating DSO's grid data, and demonstrated the application of the extracted grid data to the problem of allocating PV generation capacity in a distribution grid fairly across all the nodes.

The initial phase involves the extraction of relevant data from the DSO's grid database, which includes information on grid topology, electrical components, and their attributes. An initial validation is performed to identify and correct any obvious errors or inconsistencies in the data. This is followed by a more advanced validation procedure, which involves solving, for each grid, an offline load flow under nominal loading conditions to ensure that all grid quantities remain within physical and statutory limits, thereby verifying the overall consistency of the grid data. The results of both validation procedures are then presented to a grid expert, who cross-checks them with the DSO's internal information (e.g., maintenance reports, paper plans) to determine the necessary updates to the grid data database.

We have also presented an application of the validated grid data to compute a fair distribution of PV hosting capacity among the nodes of a distribution grid. We have formulated an optimization problem to minimize the cost of operation of the grid over the lifespan of PV plants subject to a variable degree of spatial fairness. The analysis evaluates the cost of having a fair distribution of PV capacity in the grid, irrespective of the node's location. \todo{While the method is general and can be applied to any distribution grid provided with the necessary grid information, the results for the proposed use case showed that fairness can be significantly improved with a marginal increase in costs; however, perfect fairness becomes expensive.}

The contributions of this study can inspire other DSOs to adopt similar strategies to validate their grid data and valorize them into actionable resources to make decisions that support and mitigate the impact of the energy transition on distribution grids.



\section{Acknowledgments}
This research is supported by Innosuisse in the context of the flagship project STORE (grant agreement 108.230).



\bibliographystyle{unsrtnat}
\bibliography{biblio}


\end{document}